\documentclass[fleqn,10pt]{wlscirep}

\title{Predicting knee osteoarthritis severity: comparative modeling based on patient's data and plain X-ray images}

\author[1,*]{Jaynal Abedin}
\author[2] {Joseph Antony}
\author[2]{Kevin McGuinness}
\author[2,3] {Kieran Moran}
\author[2] {Noel E O'Connor}
\author[1,5]{Dietrich Rebholz-Schuhmann}
\author[1,4] {John Newell}

\affil[1]{Insight Centre for Data Analytics, National University of Ireland Galway, Ireland}
\affil[2]{Insight Centre for Data Analytics, Dublin City University, Dublin, Ireland}
\affil[3] {School of Health and Human Performance, Dublin City University, Dublin, Ireland}
\affil[4] {School of Mathematics, Statistics and Applied Mathematics, National University of Ireland Galway, Ireland}
\affil[5] {ZB MED -- Information Centre for Life Sciences, University of Cologne, Germany}

\affil[*]{joystatru@gmail.com, jaynal.abedin@insight-centre.org}

\begin{abstract}
Knee osteoarthritis (KOA) is a disease that impairs knee function and causes pain. A radiologist reviews knee X-ray images and grades the severity level of the impairments according to the Kellgren and Lawrence grading scheme; a five-point ordinal scale (0--4). 
In this study, we used Elastic Net (EN) and Random Forests (RF) to build predictive models using patient assessment data (i.e. signs and symptoms of both knees and medication use) and a convolution neural network (CNN)  trained using X-ray images only. Linear mixed effect models (LMM) were used to model the within subject correlation between the two knees.  The root mean squared error for the CNN, EN, and RF models was $0.77$, $0.97$ and $0.94$ respectively. The LMM shows similar overall prediction accuracy as the EN regression but correctly accounted for the hierarchical structure of the data resulting in more reliable inference.  Useful explanatory variables were identified that could be used for patient monitoring before X-ray imaging. Our analyses suggest that the models trained for predicting the KOA severity levels achieve comparable results when modeling X-ray images and patient data. The subjectivity in the KL grade is still a primary concern.

\end{abstract}
\begin{document}

\flushbottom
\maketitle

\thispagestyle{empty}

\section*{Introduction}
Osteoarthritis (OA) is the result -- and the observable status -- of inflammatory processes in a joint leading to functional and anatomical impairments. 
The resulting status often shows irreversible damages to the joint cartilage and the surrounding bone structures \cite{arden2014atlas,eyre2004collagens}. 
The knees are the most commonly affected joints in the human body and knee osteoarthritis (KOA) is more prevalent in females aged 60 years or more compared to males of the same age (13\% vs 10\%)~\cite{murphy2011subgroups}. 
Severity of KOA amongst females with more than 55 years of age is higher compared to their male counterparts and the severity of KOA is higher compared to other types of OA~\cite{zhang2010epidemiology,peat2001knee}.
Approximately one in every six patients consult with a general practitioner in their first year of an OA episode ~\cite{zhang2010epidemiology,peat2001knee}. 
The incidence of KOA has a positive association with age and weight and the prevalence is more common in younger age groups, particularly those who have obesity problems~\cite{bliddal2009treatment}.

Swelling, joint pain, and stiffness are the prominent symptoms among others, such as restrictions in movement including walking, stair climbing, and bending~\cite{heidari2011knee}. 
The symptoms worsen over time and elderly patients are affected more frequently than patients in other age groups. 
The presence of OA in the knee reduces activity in daily life and eventually leads to disability, which can incur high costs related to loss in productivity~\cite{altman2010early}. 
It is estimated that functional impairment of the knee and the hip are the eleventh highest disability factors~\cite{cross2014global} contributing to considerable socio-economic burden with an estimated cost per patient per year of approximately 19,000 Euro~\cite{puig2015socio}. 
The estimated prevalence of disability due to arthritis is expected to reach 11.6 million individuals by the year 2020~\cite{centers1994arthritis}, which is greater than the estimated risk of disability attributable to cardiovascular diseases or any other medical condition~\cite{guccione1994effects}. 
Total joint replacement surgery is the most favorable option to treat advanced stage OA. 
However, diagnosing the status of KOA at an early stage and providing behavioral interventions could be beneficial for prolonging a healthy life for a patient~\cite{karsdal2016disease}.

In a review of possible risk factors of KOA, Heidari~\cite{heidari2011knee} concluded that age, obesity, gender (i.e., female), repetitive knee trauma and kneeling are the most common risk factors for KOA. 
The common symptoms include pain, functional impairment, swelling and stiffness. 
The severity of KOA and the pain status is measured based on the Kellgren and Lawrence (KL) scale of 0 to 4 by visual inspection of the knee X-ray images~\cite{kellegren1957radiological}.

Considering the impact of KOA on disability and the subsequent unavoidable economic burden, there is a need to quantify the severity of KOA  during the early stages of development. 
KOA severity level helps in determining appropriate treatment decisions and for the monitoring of  disease progression~\cite{braun2012diagnosis}. 
The classical way of quantifying KOA severity is by  inspection of X-ray  images of the knee by a radiologist who then grades the images according to the KL scale (from 0 for ``normal'' up to 4 for ``severe'' stage)~\cite{kellegren1957radiological}. 
This approach suffers from high levels of subjectivity as there is no gold standard grading system: the semi-quantitative nature of the KL grading scale creates ambiguity, thus giving rise to disagreements between raters (for details please refer to~\cite{kellegren1957radiological,gossec2008comparative,sheehy2015validity}).

To reduce the influence of subjectivity in quantifying KOA severity from X-ray images, computer-aided diagnosis has been very helpful~\cite{dacree1989automatic}. 
To date the sample size of available images has been the main limiting factor to train an efficient model~\cite{shamir2010assessment,woloszynski2012dissimilarity,shamir2009early,thomson2015automated}. 
The Osteoarthritis Initiative (OAI)~\cite{eckstein2007imaging} and the Multi-centre Osteoarthritis Study (MOST)~\cite{segal2013multicenter} mitigated this small sample size limitation by making thousands of patients data and X-ray images available. 
Recently, several researchers have used these resources to develop an automatic approach for quantifying  KOA severity by analyzing X-ray images~\cite{shamir2009early,oka2008fully,antony2016quantifying,antony2017automatic,tiulpin2018automatic}. 
Although there have been multiple attempts to quantify KOA severity based on an automated analysis of X-ray images, so far there has been no attempt to build a predictive model on a patient's assessment data such as signs, symptoms, medication and other characteristics about a patient (later on referred as patient's questionnaire data) and to compare this approach against the X-ray based prediction. 
Developing predictive models using patient data other than X-ray images offers additional advantages such as identifying those variables that contribute strongest towards predicting the severity of KOA. 
A good predictive model based on patient's questionnaire data could reduce treatment costs and could also contribute to a prolonged healthy life of a patient due to early behavioral intervention.

The Osteoarthritis Initiative (OAI) is a multi-center longitudinal study for men and women sponsored by the National Institute of Health (NIH) to better understand KOA. 
Data collected through the OAI can provide useful information about the marginal distributions of relevant patient characteristics, their demographics, signs \& symptoms and medication history. 
To date, there are more than 200 scientific publications that have used data collected through the OAI including several attempts to automate the KL grade quantification using X-ray images. 
But to date no study (or publication) has tried to predict KOA severity based on patient questionnaire data.
Our primary goal is to compare the prediction accuracy of a statistical model based on patient questionnaire data to the prediction accuracy based on X-ray image based modeling to predict KOA severity score.
In this paper we present several statistical approaches to predict the severity of KOA using patient questionnaire data.
Furthermore, we use a convolution neural network (CNN) model to predict the same outcome using corresponding X-ray images for the same patients. 
The performance of both the approaches has been compared using the calculated root mean squared error on a validation set. 
As a secondary goal, we identified key variables with the strongest predictive ability, which may be useful to monitor a patients over time and design early interventions for prolonging healthy life in patients of concern.  

\section*{Results}
\subsection*{Exploratory analysis}
The OAI dataset contains data for $4,796$ individuals. 
After initial pre-processing, $2,951$ patients with sufficient data on potential candidate explanatory variables were selected, representing 62\% of the original patients. 
The remaining 38\% of individuals did not have enough data for the potential explanatory variables and were not included in the analysis. The list of candidate variables, their labels and type (binary, numeric and categorical) has been provided in a supplementary table (Supp table 1).

To train and validate the predictive models we used a training and validation data split as shown in Table ~\ref{tab:tab1} (roughly a 70\% - 30\% split). 
To make valid comparisons, we used the same validation set in the models developed on patient's questionnaire data and the model developed using X-ray images~\cite{antony2016quantifying,antony2017automatic}. 
The validation set contained data for both knees for $846$ patients, i.e.~$1,692$ data points for a knee in total. 
The training set for the predictive models included data from $2,105$ patients, i.e.~$4,210$ knees.

\begin{table}[!ht]
\centering
\begin{tabular}{|l|r|r|r|}
\hline
Severity level & Training: Freq (\%)& Validation: Freq (\%) & Total: Freq (\%) \\ \hline
Level 0 &1818 (43.2) &685 (40.5)  &2503 (42.4) \\ \hline
Level 1 &728 (17.3)  &312 (18.4) &1040 (17.6) \\ \hline
Level 2 &1045 (24.8)  &416 (24.6) &1461 (24.8) \\ \hline
Level 3 &503 (12.5)  &237 (14.0) &740 (12.5) \\ \hline
Level 4 &115 (2.7)  &42 (2.5) &157 (2.7) \\ \hline
\end{tabular}
\caption{\label{tab:tab1}Distribution of KOA severity between training and validation data}
\end{table}

The validation data that contains 30\% of the original patients data are the same patients information that has been used in X-ray image based modeling. 
To make our results comparable with X-ray based prediction we used same validation patients information, although we performed cross validation to check sensitivity of our entire analysis. The cross validation result is consistent with our original 70-30 split. 

Relevant summary statistics for patient characteristics are given in Table~\ref{tab:tab2} for the entire dataset and for the training and the validation subsets. Good balance is evident when comparing the mean and variability for each patient characteristic across the training and validation data and it is plausible that they can be considered as representative samples taken from the same overall population. Maintaining a similar distribution of patient characteristics between the training and validation data is paramount for making reliable inference. We dropped occupation from the patient characteristics table and from subsequent analysis as this variable has more than 30\% missing data.

\begin{table}[ht]
\centering
\begin{tabular}{|l|r|r|r|}
\hline
Characteristics & Training: Mean (SD) & Validation: Mean (SD) & Total: Mean (SD) \\ \hline
Age &60.3 (9.2) &61.1 (8.9) & 60.5 (9.1)\\ \hline
Female (Freq. \%)& 1177 (56.0) & 454 (53.7) & 1631 (55.3) \\ \hline 
Height (mm) &1685.2 (93.2) &1687.3 (92.6) & 1685.8 (93.0)\\ \hline
Weight (kg) &80.7 (16.3) &80.5 (15.7) & 80.6 (16.1)\\ \hline
BMI (kg/$m^2$) & 28.3 (4.8) & 28.2 (4.6) & 28.3 (4.7)\\ \hline
Systolic &123.3 (15.9) &123.7 (16.7) & 123.3 (16.1)\\ \hline
Diastolic &75.5 (9.8) &75.4 (9.6) & 75.5 (9.8)\\ \hline
\end{tabular}
\caption{\label{tab:tab2} Summary statistics of patient characteristics between complete, training and validation data}
\end{table}

\begin{figure}[ht]
\centering
\includegraphics[width=0.8\textwidth, height=4.35 in]{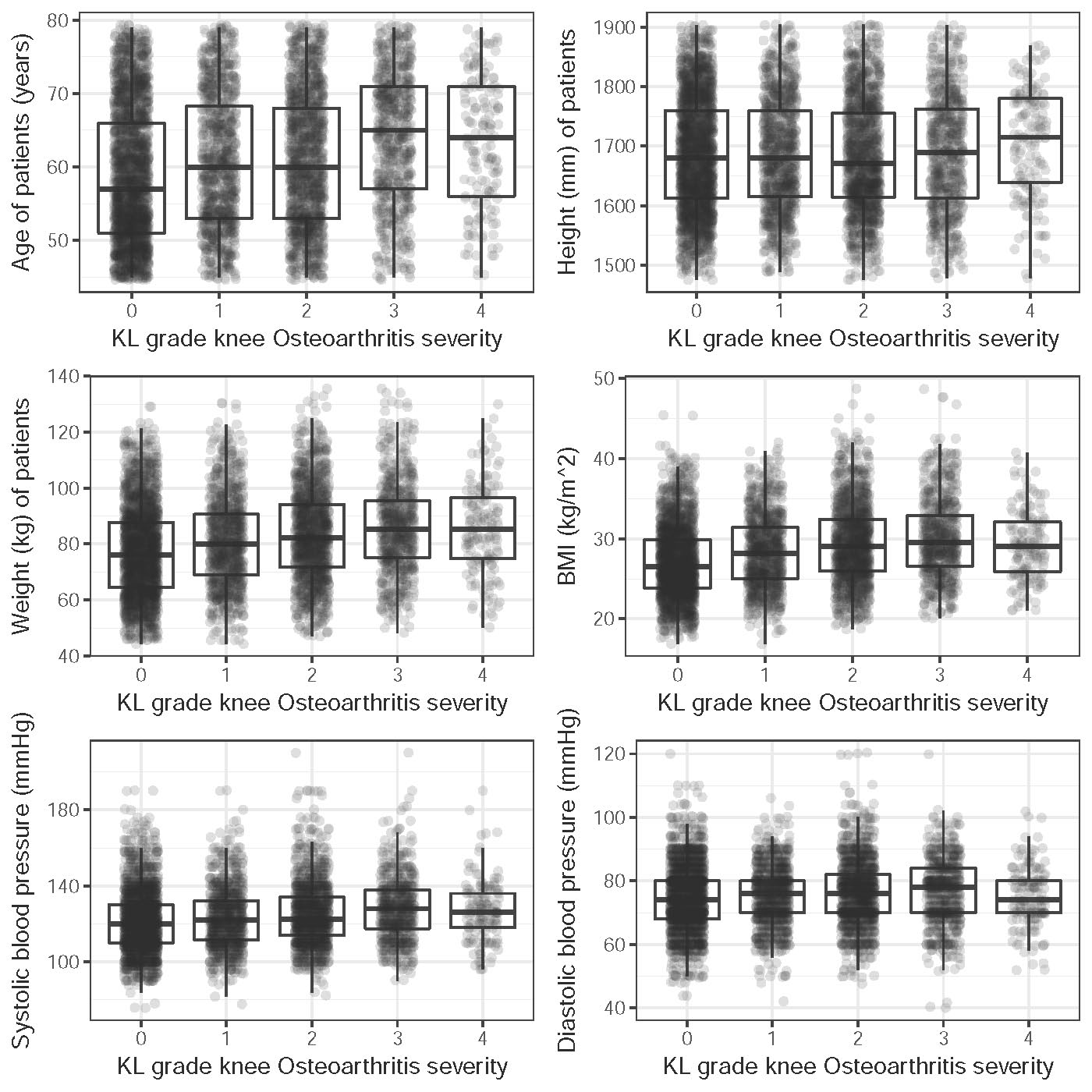}
\caption{Boxplots of patient characteristics by Knee Osteoarthritis Severity}
\label{fig:patientCharacteristics}
\end{figure}

The box-plot (Figure~\ref{fig:patientCharacteristics}) displays the distribution of several patient characteristics. We have introduced minor displacement quantity (jitters) along the horizontal axis and alpha-blending to make the display of the distribution of the points clearer. 
The level of KOA severity appears to be higher among elderly people. Height, weight and BMI show similar patterns but in contrast the distribution of blood pressure measurements does not indicate any strong obvious pattern with severity score.  (Figure~\ref{fig:patientCharacteristics}).

\begin{figure}[ht]
\centering
\includegraphics[width=0.95\textwidth]{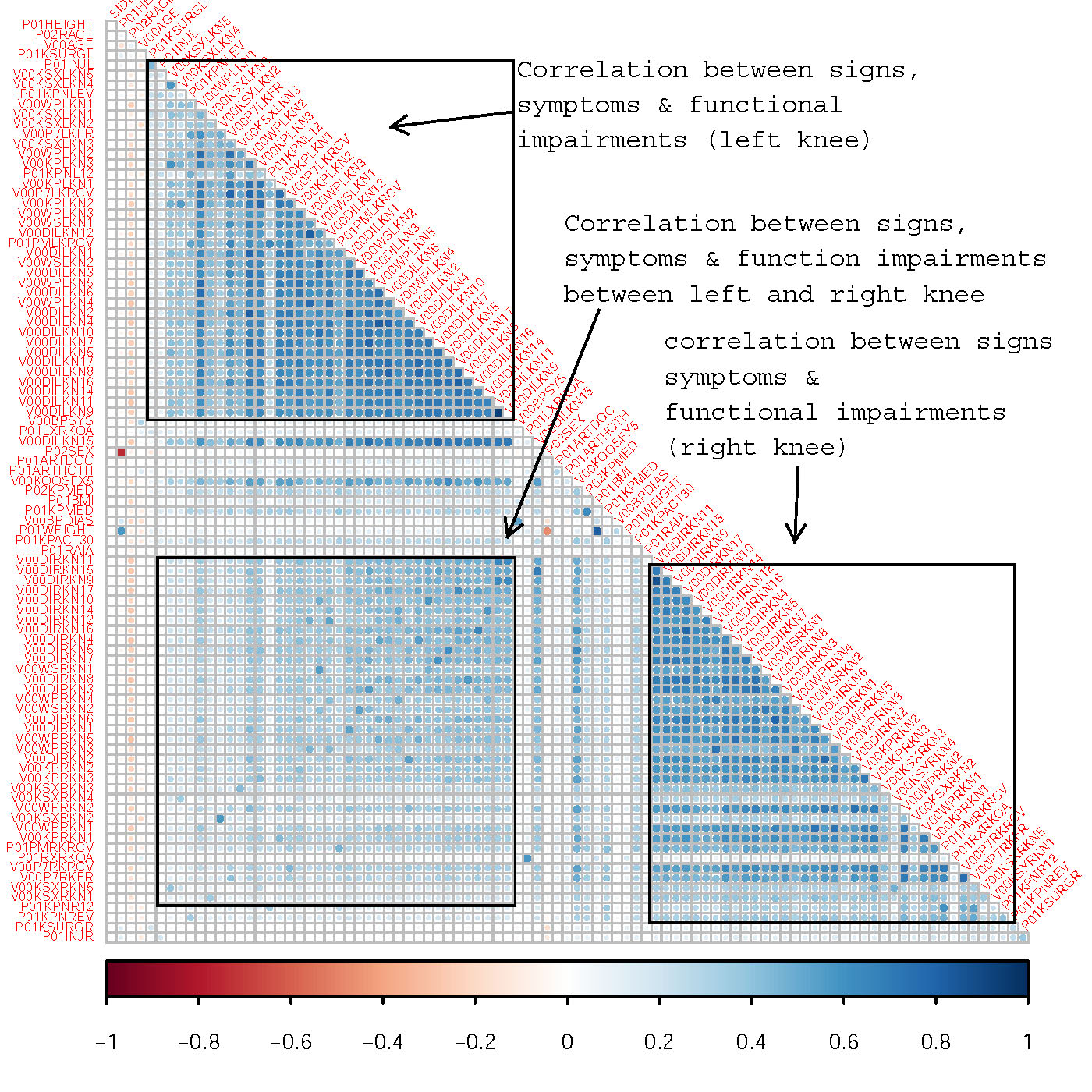}
\caption{Correlation among predictors (dark color indicates stronger correlation)}
\label{fig:predictorsCorrelation}
\end{figure}

The data contains a mixture of continuous, binary, and categorical predictors. To observe the relationship among such a collection of predictor variables we calculated Pearson correlation between continuous predictors, polyserial correlations between continuous and categorical predictors and polychoric correlation between categorical predictors~\cite{kolenikov2004use}. 
The correlation matrix (Figure~\ref{fig:predictorsCorrelation}) shows the relationship among the predictor variables of interest where a higher color intensity indicates a stronger correlation between variables. The blue color indicates a positive correlation and red color indicates a negative correlation. We can observe that the predictor variables are positively correlated with each others to a moderate degree. Patients sex, height and weight shows weak negative correlation with other variables but only sex and height show a strong negative correlation. The upper block represents correlation among signs and symptoms in the left knee whereas the lower right block represents correlation among signs and symptoms of the right knee. The lower left block is the correlation between signs and symptoms of left to right knee. Other than the three blocks of correlation there are some variables that represent neither of the knees; rather, those variables represents medication history and other characteristics (Figure~\ref{fig:predictorsCorrelation}).  What is clear from Figure~\ref{fig:predictorsCorrelation} is the large number of candidate predictors and the presence of multicollinearity amongst predictors which will have to be accounted for accordingly in any subsequent model.

Among the five levels of KOA severity there were very few patients from severity level 4 (KL grade) in comparison to other categories. Overall $42\%$ of patients were from severity level 0 indicating the normal knee followed by mild severity level 2 with $24\%$  and doubtful severity level 1 with $17\%$. The distribution of severity level frequencies across the training and the validation data is well balanced indicating that it is plausible that the training and validation data came from same underlying population (Table~\ref{tab:tab1}.)

\subsection*{Model building, evaluation, and comparison}
\begin{figure}[ht]
\centering
\includegraphics[width=0.75\textwidth, height= 4.5 in]{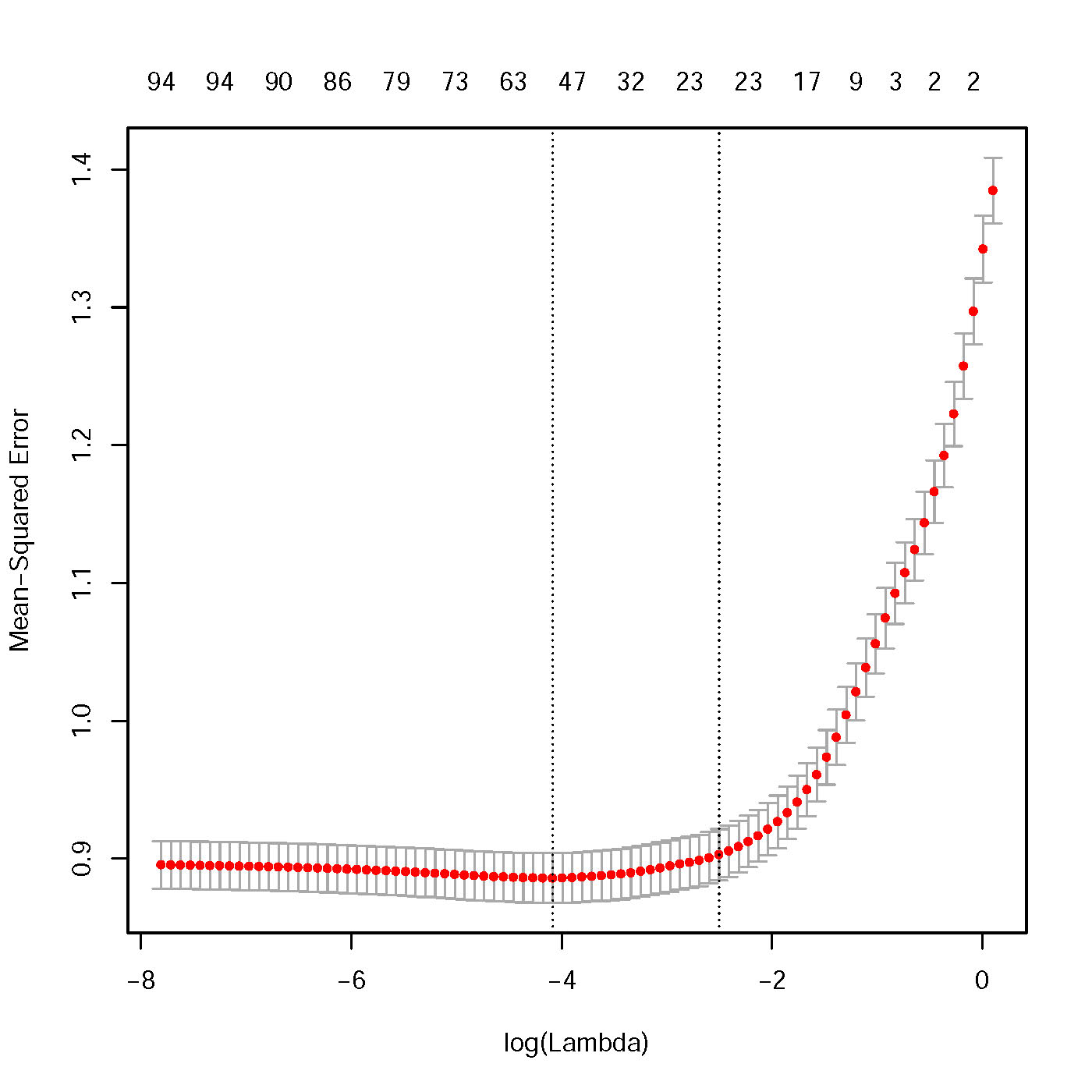}
\caption{Estimation of hyper-parameter $\lambda$. The Y-axis represents RMSE for different values of $\lambda$ whereas the upper horizontal axis represents the number of predictors. The RMSE increases as the number of predictors decrease but stabilizes after a certain number of predictors are added. The red points represents RMSE and the gray line segments represents a 95\% confidence interval corresponding to each RMSE. The optimal value of $\lambda$ is the minimum value corresponding to a steady-state RMSE.} 
\label{fig:lambdaEstimate}
\end{figure}

\begin{figure}[ht]
\centering
\includegraphics[width=0.8\textwidth]{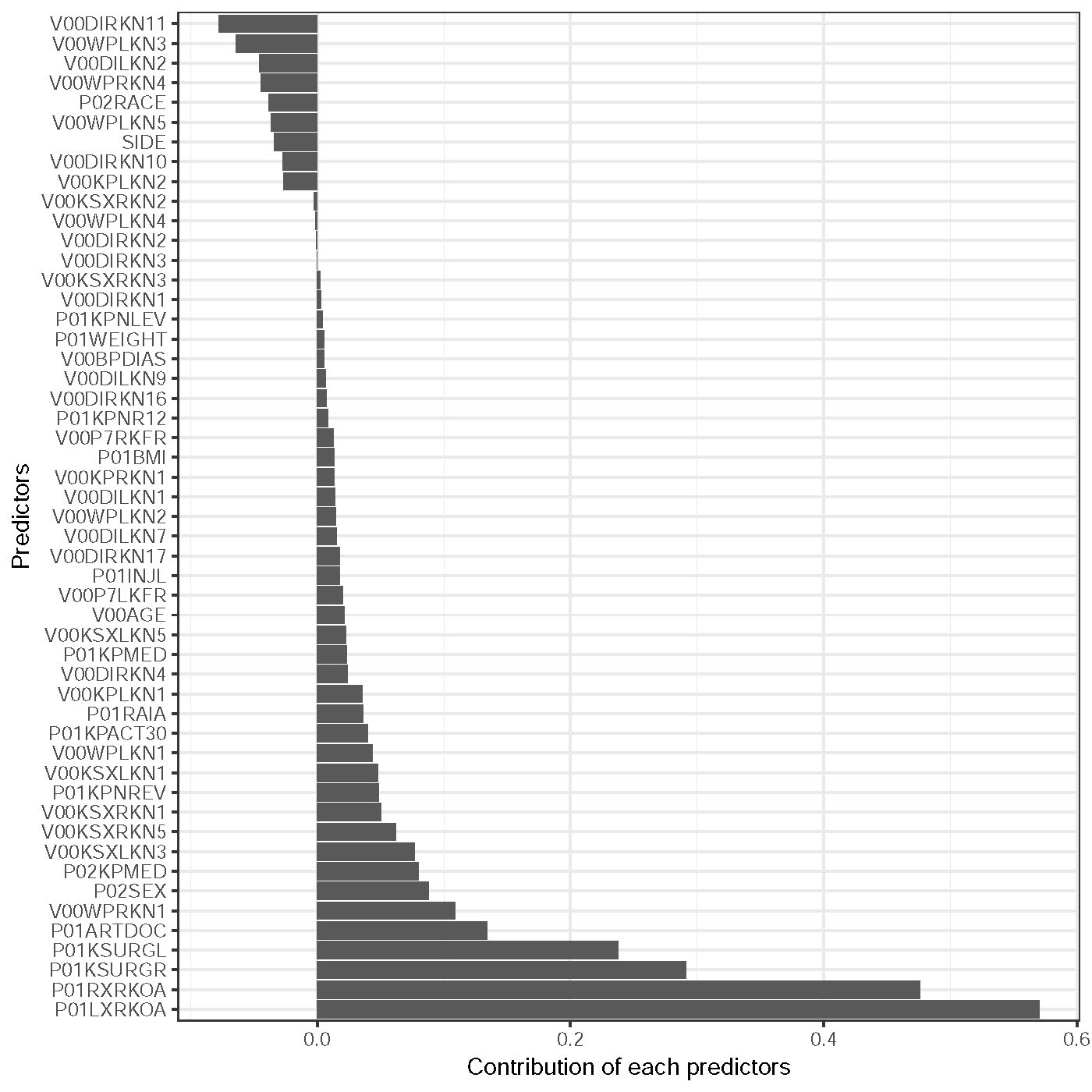}
\caption{Contribution of each variable on KOA severity score prediction by Elastic Net Regression}
\label{fig:ENcontribution}
\end{figure}

Initially an Elastic Net regression~\cite{zou2005regularization}, a weighted combination of LASSO and Ridge regression, was fitted. An Elastic Net regression model can be used to select variables with high predictive power. The weighting is controlled by the mixing parameter $\alpha$ that controls the amount of mixing between LASSO and Ridge penalties, whereas the parameter $\lambda$ controls the amount of shrinking in the regression coefficients. To estimate a suitable value for the shrinkage parameter $\lambda$ we performed repeated cross validation using a fixed $\alpha=0.5$, which corresponds to the minimum cross-validation RMSE. Using this value of $\alpha$ the value of $\lambda$ that also minimizes the RMSE (Figure~\ref{fig:lambdaEstimate}) was selected. The contribution (i.e. direction and magnitude) of each predictor variable has been extracted from the corresponding estimated regression coefficients (Figure~\ref{fig:ENcontribution}).

A Random Forest~\cite{breiman2001random} regression model was then fitted using differing numbers of trees where the RMSE was calculated for each scenario. Based on these evaluations, we found that using 100 trees produced the lowest RMSE in the validation set. We also identified those predictors with highest variable importance in terms of improved predictive ability of the final forest.

The overall RMSE for the Elastic Net regression model is $0.97$ and the RMSE for the random forest model is $0.94$. Both models give higher accuracy for the prediction of the severity levels 1 and 2 in contrast to the other categorical levels. The RMSE from the X-ray image based CNN model is $0.77$, which is slightly lower than the RMSE from the Elastic Net regression and the Random Forest model. The advantage of using Elastic Net regression over a Random Forest regression model and X-ray image based CNN model is that we can easily identify the variables that have high predictive power and also the direction of the contribution of each variables by looking at the magnitude and sign of their regression coefficients (Figure~\ref{fig:ENcontribution}).

The Elastic Net regression model produced higher prediction accuracy for severity level 1 and 2, in comparison to other levels. A similar result is noted in the predictions by the Random Forest model. The overall RMSE of Elastic Net regression and Random Forest regression models are $0.974$ and $0.943$. The overall accuracy of the CNN model is higher than Elastic Net and Random Forest regression. The performance of each of the three models show their lowest outcome for the KOA severity level 4 as there is less data available in that category. Relatively higher accuracy in predicting KOA severity using an X-ray based CNN model has been observed however, the margin of difference between the RMSE of the predictions from the X-ray image based CNN model in comparison to the predictions from the patient's questionnaire data models is considerably small. Table~\ref{tab:tab3} shows the RMSE for the models trained with patient data (Elastic Net and Random Forest)and the model trained with X-ray images (CNN regression).

\begin{table}[!ht]
\centering
\begin{tabular}{|l|r|r|r|r|}
\hline
Severity Level & Elastic Net Regression & Linear Mixed Model (LMM)& Random Forest Regression & CNN Regression \\ \hline
Level 0 &0.917  &0.920  & 0.909    &  0.816  \\ \hline
Level 1 & 0.563 &0.591  &  0.511  &  0.485 \\ \hline
Level 2 & 0.881 &0.895  &  0.853  & 0.840\\ \hline
Level 3 &1.320  &1.320  & 1.270   & 0.795 \\ \hline
Level 4 &2.140  &2.10   & 2.02   & 0.846\\ \hline
{\bf Overall} & 0.973   & 0.978  & 0.943 & 0.770\\ \hline
\end{tabular}
\caption{\label{tab:tab3} Estimated RMSE from different models for each level of KOA severity level}
\end{table}

Both the Elastic Net and Random Forest models allow us to variable importance of the individual predictors on overall predictive ability. 
There are some variables commonly identified by both these models with higher contribution towards the final predictions. However, the variables identified by the Elastic Net have more interpretable properties than the variables selected by the Random Forest model. 
The sign of the regression coefficients in an Elastic Net regression allows us to understand the direction of the contribution; whether it increases the severity score or reduces it. 
A negative sign indicates a reduction in the overall severity score for increasing values of the predictor, whereas a positive sign indicates an increase in the severity score for increasing values of the predictor. 
The direction of the contribution by predictors selected by the random forest model is unclear as it gives similar importance to both directions. 
Figure~\ref{fig:ENcontribution} shows the sign and magnitude of the contribution for each of the selected variables. 
The positive sign indicates an increase in the severity score whereas a negative sign indicates a decrease in the severity score. 
The identified variables could be a proxy indicators of patient knee's anatomical structure which ultimately indicates the level of severity.

Since the data have a hierarchical structure (i.e. knee nested within patient) the number of replicates at the individual level is more appropriately modeled using a mixed effects model with a random effect to capture the correlation between knees within an individual.  
To explore the random effect of patient level information, a linear mixed effect model was fitted using the predictor variables initially selected from the Elastic Net regression. 
There is clear evidence for the need of random effects due to the study design.  In addition a small p-value (less than 0.001) was evident for the test for the need of the random effect term due to subject level within knee correlation in the model. 
The intra-class correlation coefficient is 0.265 which indicates that the proportion of the variance explained by the random effect component (patient level information) in the population 26.5\%. 
The overall RMSE for the linear mixed effects model is 0.978, which is almost the same as the RMSE from the Elastic Net regression. However, the predicted severity levels, and more importantly the corresponding uncertainty, is correctly adjusted to account for the within patient correlation.

\section*{Discussion}

Judging the impairment for patients with KOA requires a thorough understanding of the disease condition. Expert radiologists or clinicians assess the functional knee impairments and the KOA severity level from the X-ray images. Ideally, the image analysis should give an objective measure of the impairments; however, in reality not all functional impairments show up in anatomical transformations of the knee, and the patho-physiological evaluation relies on the subjective perception of the patient and the physician jointly.

Our primary goal was to explore whether the prediction accuracy of a statistical model based on
patient’s questionnaire data is comparable to the prediction accuracy based on X-ray image based modeling to predict KOA severity.
We have demonstrated that statistical models, using patients' questionnaire data, could predict KOA severity level with a good level of accuracy (RMSE: 0.974 \& 0.943). 
The prediction performance of the statistical models presented in this paper are comparable to models using X-ray image data based on model performance as assessed by RMSE measures~\cite{antony2016quantifying, antony2017automatic, tiulpin2018automatic}. 
In particular we have demonstrated that functional impairment at severity levels 1 and 2 can be predicted by our statistical models (Elastic Net \& Random Forest and LMM) trained from the patients' assessment data to a level of accuracy similar to the accuracy achieved on the basis of CNN model trained on X-ray images. 
There are very subtle structural variations in the knee joints (minimal joint space narrowing (JSN) and osteophytes formation) belonging to grade 0 and grade 1, and these are not fully reflected in the KL grades. Also, there are relatively large overlaps in the JSN measurements for KL grades 0 and 1 compared to the other grades~\cite{hart2003kellgren}. These factors make them challenging to distinguish by inspecting the X-ray images. Also, patients share almost similar distribution on their characteristics, signs, symptoms and functional impairments. Due to very subtle differences of the predictors between KOA levels the prediction accuracy gets affected.

We were able to identify the key variables that contributed most to the predictive ability in our models. These identified variables can be monitored over time to assess the progression of KOA severity. The strong indicator variables are reporting on knee baseline radiographic OA status for the right or left knee (P01LXRKOA, P01RXRKOA) and on treatments such as surgery on the right or left knee (P01KSURGR, P01KSURGL) as well as other reasons to see the doctor (P01ARTDOC). Patient's sex also plays important role in predicting KOA severity. The next indicator variables cover medication (P02KPMED) and functional impairments, pain or other symptoms to the right or left knee (V00WPRKN1, V00KSXLKN3, V00KSXRKN5, V00KSXRKN1, V00KSXLKN1, V00WPLKN1, P01KPNREV, P01KPACT30). A final parameter notes whether a doctor ``ever said you have rheumatoid arthritis or other inflammatory arthritis'' (P01RAIA). The predictors variable that we found as important predictor of KOA severity were also reported important risk factor in previous studies~\cite{heidari2011knee, hunter2008symptoms}.

Importantly, an early behavioral intervention could be developed based on the identified variables to prolong the healthy life of a patient. By observing the identified variables that have higher predictive ability to predict KOA severity, we can identify the subjects who are currently taking medication for pain relief and facing functional difficulty in their daily life. Variables representing limited knee functions in particular are the potential indicators for quantifying KOA severity that could lead to developing targeted interventions for further treatment and medications.

When making predictions the LMM is favored as it is the only approach that correctly adjusts for the hierarchical structure present in the data.  It is interesting that the severity levels 1 and 2 can be predicted with good accuracy in all the four models (EN, RF, LMM, and CNN), while the other levels of severity are more challenging to predict. For higher severity levels, i.e. levels 3 and 4, this could be due to the lack of patient data, i.e. the sample sizes at these levels are smaller than for levels 1 and 2 (Figure~\ref{fig:patientCharacteristics} \& table~\ref{tab:tab2}).

As a conclusion based on the results in this paper, we can say that the patients' questionnaire data can predict KOA severity level with good accuracy and it is comparable with the prediction based on X-ray images. Patient's assessment data also enables us to identify some of the key variables that can be used to design early interventions and monitor the patients over the treatment period. The accuracy of the model developed using patient's assessment data is almost comparable to the CNN model. Moreover, the statistical models have an edge over the CNN model by identifying key variables that helps the physicians to design interventions and helps the patients for further treatment.

There is at least one potential limitation in developing statistical models to predict KOA severity, that is the KL grade score itself is not a gold standard and suffers from subjectivity. The KL grade is dependent on the perception of the radiologist who is inspecting the X-ray images. In the model building process, we are effectively using a quasi-gold standard outcome. Considering this potential limitation, one way to improve the prediction accuracy could be to build a model of the X-ray image data in combination with the patients' assessment data. 
The prediction of KOA severity based on patients data shows comparable accuracy, it would be interesting to see the performance of prediction based on a statistical model combining both patient's questionnaire data and with X-ray images. 

\section*{Methods}
\subsection*{Data}
The data used in this study were obtained from the Osteoarthritis Initiative (OAI), which is available for public access at \url{http://www.oai.ucsf.edu/}. 
The specific dataset used is labeled 0.2.2. This is the data from the multi-center longitudinal and prospective observational study of KOA. We have used the baseline dataset for this work. The description of each variable used in our analysis has been given in the supplementary table 1.

\subsection*{Data pre-processing and descriptive statistics}
\begin{figure}[ht]
\centering
\includegraphics[width=\linewidth, height=1.65 in]{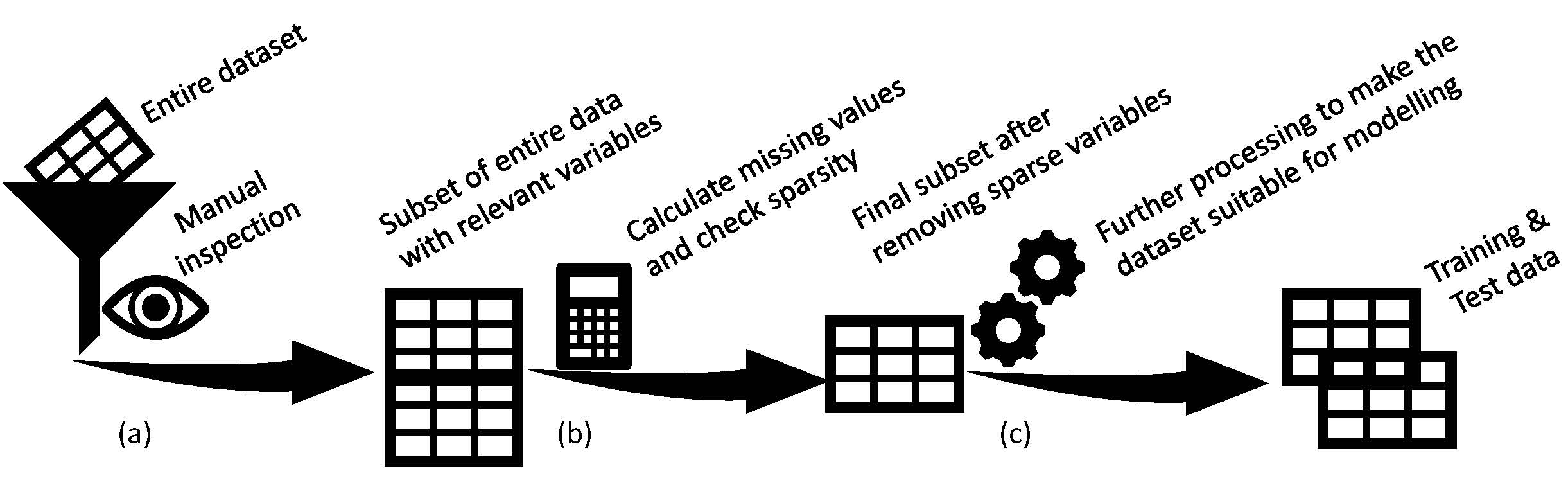}
\caption{Data pre-processing work flow: (a) Inspecting entire dataset manually to get subset of relevant candidate variables, (b) calculate percentage of missing values for each variables and also inspect sparsity of the categorical variables. Drop a variable that has more than 15\% missing values or very low e.g. less than 5\% into one category in a binary variable, (c) creating dummy variables from multi-category variables and then split the dataset into training and test data for predictive model building.}
\label{fig:dataProcessingSteps}
\end{figure}

The baseline dataset contains a large number of variables related to patients' characteristics, their vital signs, symptoms of KOA, medication history, and functional impairment. At the early stage of the analysis, we manually inspected each of the variables and selected a subset of candidate variables that were clinically relevant and previously reported risk factor for KOA~\cite{heidari2011knee,hunter2008symptoms}. We inspected the completeness of the data in terms of missing values. We calculated the amount of missing values (in percent) for each variable. The variables that had at least 85\% non-missing values were kept for further analysis.  If it can be assumed that missing data are missing at random (i.e. missingness is explained by the covariates available) a multiple imputation step is unnecessary if a linear mixed model is used as the likelihood is correctly specified under this assumption.  Moreover, we excluded categorical variables with very low discriminatory power, for example, the variables with very low frequency in one of the categories compared to the rest within the same variable. The reduced set of candidate variables was used for further processing and analysis. 
The dataset was split into two parts, training and validation sets, by taking a random sample of 70\% of the data for training and the remaining 30\% for validation. To make valid comparisons, we used the same validation set in the models developed on patient's questionnaire data and the model developed using X-ray images~\cite{antony2016quantifying,antony2017automatic}.
The data pre-processing steps are summarized in (Figure \ref{fig:dataProcessingSteps}). 

To summarize and explore the explanatory variables, we calculated descriptive statistics: mean and standard deviation for numeric variables, frequency and percentage for categorical variables. The relationship among predictors was also explored by calculating Pearson correlation between numeric variables, polyserial correlations between numeric and categorical variables and polychoric correlation between categorical variables~\cite{kolenikov2004use}.  

The KL grade score was recorded on an ordinal scale from 0: normal to 4: severe. To model an ordinal outcome, ordinal logistic regression~\cite{mccullagh1980regression, anderson1984regression} is the typical approach used. We fitted ordinal logistic regression models, but the prediction performance was poor. In this paper we have treated the severity score as a continuous response to investigate if this would improve predictive ability.   
Moreover, the data are hierarchical in structure; for each patient we have data for both knees. To capture this structure appropriately we have used a linear mixed effect model incorporating a random effect at the subject level~\cite{laird1982random}.

\subsection*{Elastic Net regression}
Elastic Net regression is a combination of ridge regression and LASSO, and this model is appropriate in the presence of correlated predictors~\cite{zou2005regularization}. 
We denote the outcome variable: KL grade score by $Y$ (considered as a continuous variable) and all predictors by $X_1, X_2,...,X_p$. The Elastic Net regression linearly combines $L_1$ and $L_2$ penalties as follows:
\begin{eqnarray}
\label{eq:schemeP}
	SSE = \sum_{i=1}^n (y_i - \hat{y_i})^2+\lambda [(1-\alpha) \sum_{j=1}^p \theta_j^2 + \alpha \sum_{j=1}^p |\theta_j|]
\end{eqnarray}
The $L_1$ penalty is defined as the sum of absolute value of the regression coefficients and the $L_1 = \sum_{j=1}^p |\theta_j|$ and $L_2$ penalty is defined as the sum of squared values of regression coefficients: $L_2 = \sum_{j=1}^p \theta_j^2$. The amount of mixing between two penalty term is controlled by a mixing parameter $\alpha$. If the value of $\alpha =0$ then it leads to a ridge regression whereas a value of $\alpha=1$ leads to LASSO regression. The hyper-parameter $\lambda$ controls the amount of shrinkage of regression coefficients for various values of $\alpha$. A higher value of $\lambda$ leads to shrink the regression coefficients towards zero and a very small value of $\lambda$ has little effect on the regression coefficients. Using both $L_1$ and $L_2$ penalty enables us to select appropriate variables that have higher predictive power by shrinking some of the regression coefficient to zero using an appropriate value of hyper-parameter $\lambda$. To estimate the most suitable value for the shrinkage parameter $\lambda$ we performed repeated cross validations with fixed values of the mixing parameter $\alpha=0.5$ and choose the value of $\lambda$ that minimizes the root mean squared error (RMSE). Figure~\ref{fig:lambdaEstimate}, shows the cross-validation results while selecting $\lambda$.  

\subsection*{Random Forest}
Random Forests (RF) are an ensemble method that combines the predictive ability of multiple tree based models. The RF model is an extension of the original work of Tin Kam Ho~\cite{ho1995random} who developed the algorithm for random decision forests. Leo Breiman~\cite{breiman2001random} used the idea of bagging (bootstrap aggregating) and random variable selection. The principle of random forest is to combine multiple tree based models to form a single model that can achieve better accuracy compared to its individual counterparts. This method takes a random sample with replacement from the original data, then builds a decision tree model based on a random selection of variables at each branch in the tree. This process is repeated for multiple trees and stores the prediction from each tree. The predicted value is then the mode (for a categorical response) for the mean (for a continuous response) across the forest. The random forest model is popular because it can reduce the variance of single tree models and also overcomes the problem of correlated predictors as it takes only a subset of candidate predictor variables in each of the individual trees. 

\subsection*{Linear Mixed Effect Model}
There is a clear hierarchical structure in the dataset as we have patient level data along with knee level data. A linear mixed effect model (LMM)~\cite{laird1982random} is an extension of a linear model that accounts for the hierarchical structure in data.  The primary benefit of using a LMM in this paper is that the uncertainty in knee level prediction is now correctly adjusted for through the introduction of a suitable random effect.  This approach will  account for measurements on both knees collected for each subject correctly. The Intra-class correlation has been reported that indicates how much variation in the dependent variable is due to random effect component in the LMM model. A random effect model can be formulated as: 
\begin{eqnarray}
\label{eq:schemeP1}
	y_{ij} = x_{ij}^t \beta + u_{ij}^t \gamma_i + \epsilon_{ij}; i = 1, 2, \cdots, m; j = 1, 2, \cdots, n_i
\end{eqnarray}
Here $y_{ij}$ is the KL grade of i-th knee of j-th patient, $x_{ij}$ the covariate of vector of j-th member of cluster $i$ for fixed effects; $u_{ij}$ covariate vector of j-th member of cluster $i$ for random effects; $\gamma_i$ is the random effect parameter, $m$ is the the number of cluster (in our case $m=2$ representing left and right knee), $\beta$ is the regression coefficient of the fixed effect covariates.  

\subsection*{Convolution Neural Network}
In the machine learning based approach to automatically assess the KOA severity, the first step is to localize the region of interest (ROI), that is to detect and extract the knee joint regions from the X-ray images, and the next step is to classify the localized knee joints based on KL grades. In our previous study\cite{antony2017automatic}, we introduced a fully convolutional neural network (FCN) to automatically detect and extract the knee joints, and trained CNNs from scratch to predict the KOA in both discrete and continuous scales using classification and regression respectively\cite{antony2016quantifying,antony2017automatic}. 
We used the baseline X-ray images from the OAI dataset to train the CNN model. After testing different configurations, the network in Table \ref{tab:bestClsf} was found to be the best for classifying knee images. The network contains five layers of learned weights: four convolutional layers and a fully connected layer. Each convolutional layer in the network is followed by batch normalization and a ReLU activation layer. After each convolutional stage there is a max pooling layer. The final pooling layer (maxPool4) is followed by a fully connected layer (fc5) with output shape of 1024 and a softmax dense (fc6) layer with output shape of 5 representing five level of KOA severity. To avoid overfitting, a drop out layer with a drop out ratio of 0.25 is included after the last convolutional (conv4) layer and a drop out layer with a drop out ratio of 0.5 after the fully connected layer (fc5). Also, a L2-norm weight regularization penalty of 0.01 is applied in the last two convolutional layers (conv3 and conv4) and the fully connected layer (fc5). Applying a regularization penalty to other layers increases the training time whilst not introducing significant variation in the learning curves. The network is trained to minimize categorical cross-entropy loss using the Adam optimizer with default parameters: initial learning rate $(\alpha) = 0.001$, $\beta_{1} = 0.9$, $\beta_{2} = 0.999$, $\epsilon = 1\mathrm{e}^{-8}$. The inputs to the network are knee images of size 200$\times$300. This size is selected to approximately preserve the aspect ratio  based on the mean aspect ratio (1.6) of all the extracted knee joints.

\begin{table}[ht]
\centering
\begin{tabular}{l c c c c}
\toprule 
Layer & Kernels & Kernel Size & Strides & Output shape\tabularnewline
\midrule
conv1 & 32 & 11$\times$11 & 2 & 32$\times$100$\times$150 \tabularnewline
maxPool1 & -- & 3$\times$3 & 2 & 32$\times$49$\times$74 \tabularnewline
conv2 & 64 & 5$\times$5 & 1 & 64$\times$49$\times$74 \tabularnewline
maxPool2 & -- & 3$\times$3 & 2 & 64$\times$24$\times$36 \tabularnewline
conv3 & 96 & 3$\times$3 & 1 & 96$\times$24$\times$36  \tabularnewline
maxPool3 & -- & 3$\times$3 & 2 & 96$\times$11$\times$17 \tabularnewline
conv4 & 128 & 3$\times$3 & 1 & 128$\times$11$\times$17  \tabularnewline
maxPool4 & -- & 3$\times$3 & 2 & 128$\times$5$\times$8 \tabularnewline
\bottomrule
\end{tabular}
\caption{\label{tab:bestClsf} Best performing CNN for classifying the knee images}
\end{table}

\begin{figure}[ht]
\centering
\includegraphics[width=0.8\textwidth, height=1.75 in]{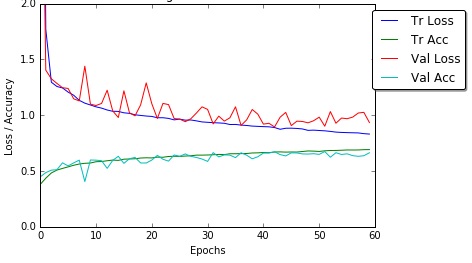}
\caption{Learning curves: training and validation losses, and accuracy of the fully trained CNN.} 
\label{fig:Lc_bestClsf}
\end{figure}

After training, this network achieves an overall root mean-squared error 0.771 on the test data. Figure \ref{fig:Lc_bestClsf} shows the learning curves whilst training this network. The learning curves show proper convergence of the training and validation losses with consistent increase in the training and validation accuracy until they reach constant values.

\bibliography{bibKOAS}

\section*{Acknowledgements}

This publication has emanated from research supported in part by a research grant from Science Foundation Ireland (SFI) under Grant Number SFI/12/RC/2289, co-funded by the European Regional Development Fund.

The OAI is a public-private partnership comprised of five contracts (N01-AR-2-2258; N01-AR-2-2259; N01-AR-2- 2260; N01-AR-2-2261; N01-AR-2-2262) funded by the National Institutes of Health, a branch of the Department of Health and Human Services, and conducted by the OAI Study Investigators. Private funding partners include Merck Research Laboratories; Novartis Pharmaceuticals Corporation, GlaxoSmithKline; and Pfizer, Inc. Private sector funding for the OAI is managed by the Foundation for the National Institutes of Health.

\section*{Author contributions statement}
Jaynal Abedin [JA] has done the overall data management, statistical analysis \& manuscript writing and coordinated with all the co-authors. Joseph Antony [JA] acquired the OAI dataset, trained the CNN model with X-ray images and contributed in writing the manuscript. Kevin McGuinness [KMG], Kieran Moran [KM] \& Noel E O'Connor [NO] critically reviewed and provided feedback to revise the manuscript, and supervised this work. Dietrich Rebholz-Schuhmann [DRS] reviewed initial candidate variables for clinical relevance and reviewed the manuscript to provide critical inputs to improve the quality. John Newell [JN] oversaw the statistical analysis and critically reviewed and suggested necessary improvements from the statistical point of view, and he also supervised the first author throughout this work. 

\section*{Additional information}
\textbf{Competing Interests:} The authors declare that they have no competing interests either financially and non-financially.

\end{document}